\documentclass[12pt]{article}
\usepackage{amsmath}
\usepackage{amssymb}
\usepackage{amsfonts}
\usepackage{graphics}

\renewcommand{\theequation}{\arabic{section}.\arabic{equation}}
\newcommand{\field}[1]{\mathbb{#1}}
\newcommand{\R}{\field{R}}
\DeclareMathOperator{\Res}{Res}

\title{Families of quasi-exactly solvable extensions of the quantum oscillator in curved spaces}
\author{C. Quesne\thanks{Electronic mail: cquesne@ulb.ac.be}\\ 
{\small\sl Physique Nucl\'eaire Th\'eorique et Physique Math\'ematique,  Universit\'e Libre de Bruxelles,} \\ 
{\small\sl Campus de la Plaine CP229, Boulevard~du Triomphe, B-1050 Brussels, Belgium}}
\date{ }
\begin{document}
\baselineskip=22pt plus 1pt minus 1pt
\maketitle
\begin{abstract}
We introduce two new families of quasi-exactly solvable (QES) extensions of the oscillator in a $d$-dimensional constant-curvature space. For the first three members of each family, we obtain closed-form expressions of the energies and wavefunctions for some allowed values of the potential parameters using the Bethe ansatz method. We prove that the first member of each family has a hidden sl(2,$\R$) symmetry and is connected with a QES equation of the first or second type, respectively. One-dimensional results are also derived from the $d$-dimensional ones with $d \ge 2$, thereby getting QES extensions of the Mathews-Lakshmanan nonlinear oscillator. 
\end{abstract}


%
%
\newpage
\section{INTRODUCTION}

During many years, there has been a continuing interest in the classical nonlinear oscillator introduced by Mathews and Lakshmanan \cite{mathews} as a one-dimensional analogue of some quantum field theoretical models. Such a nonlinear oscillator is indeed an interesting example of a system endowed with a position-dependent mass and having periodic solutions with an amplitude-dependent frequency. Since the solutions of its quantum version are also well known \cite{carinena04a, carinena07a, schulze}, it is amenable to applications in many areas of physics.\par
%
%
Its two-dimensional (and more generally $d$-dimensional) classical generalization was introduced by Cari\~ nena, Ra\~ nada, Santander, and Senthilvelan \cite{carinena04b}, who established that the nonlinearity parameter $\lambda$, entering the definitions of the potential and of the position-dependent mass, can be interpreted as $-\kappa$, where $\kappa$ is the curvature of the space, so that their model actually describes a harmonic oscillator on the sphere (for $\lambda = -\kappa <0$) or in a hyperbolic space (for $\lambda = -\kappa > 0$). The corresponding quantum model was also exactly solved in two \cite{carinena07b, carinena07c, cq15}, three \cite{carinena12}, and $d$ \cite{cq16} dimensions. It is worth observing that the oscillator in a spherical geometry had already been studied from a Lie algebraic viewpoint more than forty years ago and is known as the Higgs oscillator \cite{higgs, leemon}.\par
%
%
In a recent work \cite{cq16}, some rational extensions of the quantum oscillator in a $d$-dimensional space of constant curvature were constructed. These are also exactly solvable problems, whose bound-state wavefunctions can be written in terms of exceptional orthogonal polynomials (see, e.g., Ref.~\cite{gomez} and references quoted therein), instead of classical orthogonal polynomials for the oscillator alone.\par
%
%
Here, we plan to construct other types of extensions leading to quasi-exactly solvable (QES) Schr\"odinger equations. The latter occupy an intermediate place between exactly solvable and non-solvable ones in the sense that only a finite number of eigenstates can be found explicitly by algebraic means, while the remaining ones remain unknown. The simplest QES problems, discovered in the 1980s, are characterized by a hidden sl(2,$\R$) algebraic structure \cite{turbiner87, turbiner88, ushveridze, gonzalez, turbiner16} and are connected with polynomial solutions of the Heun equation \cite{ronveaux}. Generalizations of this equation are related through their polynomial solutions to more complicated QES problems. In such cases, to avoid dealing with high-order recursion relations, it is easier to resort to the functional Bethe ansatz method \cite{gaudin, ho, zhang}, which has proven very effective in such a context \cite{agboola12, agboola13, agboola14}.\par
%
%
Our purpose is to consider two families of quantum systems with respective potentials $V^{(1)}_m(r)$ and $V^{(2)}_m(r)$, $m=1, 2, 3, \ldots$, $0 < r < r_{\rm max}$, generalizing the oscillator potential $V_0(r) = V^{(1)}_0(r) = V^{(2)}_0(r)$ in a $d$-dimensional constant-curvature space, and to obtain exact solutions for them by means of the Bethe ansatz method. Via some small changes, we will also show that our results are applicable to one-dimensional potentials $V^{(1)}_m(x)$ and $V^{(2)}_m(x)$, $m=1, 2, 3, \ldots$, $-x_{\rm max} < x < x_{\rm max}$, extending the Mathews-Lakshmanan nonlinear oscillator $V_0(x)$.\par
%
%
This paper is organized as follows. In Section II, we review the quantum problem of the oscillator in a $d$-dimensional constant-curvature space and its bound-state energies and wavefunctions, as well as their one-dimensional counterparts. In Sections III and IV, the two families of potentials $V^{(1)}_m$ and $V^{(2)}_m$ are defined and closed-form solutions are obtained for their first three members. Section V contains the conclusion.\par
%
%
\section{THE OSCILLATOR IN A CONSTANT-CURVATURE SPACE}

In units wherein $\hbar = 2m = 1$, the quantum version of the one-dimensional Mathews-Lakshmanan oscillator is described by the Hamiltonian \cite{carinena04a, carinena07a}
\begin{equation}
  \hat{H} = - (1+\lambda x^2) \frac{d^2}{dx^2} - \lambda x \frac{d}{dx} + V_0(x), \qquad V_0(x) = \frac{\beta (\beta +
  \lambda)x^2}{1+\lambda x^2},  \label{eq:H-1} 
\end{equation}
where $\beta$ plays the role of the frequency $\omega$ in the standard oscillator and the nonlinearity parameter $\lambda \ne 0$ enters both the potential energy term $V_0(x)$ and the kinetic energy one, giving rise there to a position-dependent mass. According to whether $\lambda>0$ or $\lambda<0$, the range of the coordinate $x$ is $(-\infty,+\infty)$ or $\left(-1/\sqrt{|\lambda|}, 1/\sqrt{|\lambda|}\right)$. Such a Hamiltonian is formally self-adjoint with respect to the measure $d\mu = (1+\lambda x^2)^{-1/2} dx$.\par
%
%
The corresponding Schr\"odinger equation
\begin{equation}
  \left(-(1+\lambda x^2) \frac{d^2}{dx^2} - \lambda x \frac{d}{dx} + \frac{\beta(\beta+\lambda)x^2}{1+\lambda x^2}
  \right) \psi(x) = E \psi(x)  \label{eq:SE-1}
\end{equation}
is exactly solvable and its bound-state wavefunctions can be expressed in terms of Gegenbauer polynomials as \cite{schulze}
\begin{equation}
  \psi_n(x) \propto 
    \begin{cases}
      (1+\lambda x^2)^{-\beta/(2\lambda)} C_n^{(-\beta/\lambda)}({\rm i}\sqrt{\lambda}\, x) & \text{if $\lambda>0$,}
         \\[0.2cm]  
      (1-|\lambda| x^2)^{\beta/(2|\lambda|)} C_n^{(\beta/|\lambda|)}(\sqrt{|\lambda|}\, x) & \text{if $\lambda<0$,}   
    \end{cases}  \label{eq:wf-1}
\end{equation}
with corresponding energy eigenvalues
\begin{equation}
  E_n = \beta(2n+1) - \lambda n^2.
\end{equation}
Here the range of $n$ values is determined by the normalizability of $\psi_n(x)$ on the appropriate interval with respect to the measure $d\mu$. It is given by
\begin{equation}
  n =
    \begin{cases}
      0, 1, 2, \ldots, n_{\rm max}, \quad \frac{\beta}{\lambda}-1 \le n_{\rm max} < \frac{\beta}{\lambda},
        & \text{if $\lambda>0$,} \\[0.2cm]
      0, 1, 2, \ldots & \text{if $\lambda<0$.}
    \end{cases}  \label{eq:n-1}
\end{equation}
\par
%
%
{}For $d\ge 2$, the $d$-dimensional generalization of Hamiltonian (\ref{eq:H-1}) can be written as \cite{carinena07b, carinena07c, cq15, carinena12, cq16}
\begin{equation}
  \hat{H} = - (1+\lambda r^2) \hat{\Delta} - \lambda r \frac{\partial}{\partial r} - \lambda \hat{J}^2 + V_0(r), \qquad
  V_0(r) = \frac{\beta(\beta+\lambda)r^2}{1+\lambda r^2},  \label{eq:H-d}
\end{equation}
where $\hat{\Delta}$ denotes the Laplacian in a $d$-dimensional Euclidean space, $r^2 \equiv \sum_i x_i^2$, $\hat{J}^2 \equiv \sum_{i<j} \hat{J}_{ij}^2$, $\hat{J}_{ij}$ is an angular momentum component, and $i$, $j$ run over $1, 2, \ldots, d$. The corresponding Schr\"odinger equation is separable in hyperspherical coordinates and gives rise to the radial equation
\begin{equation}
  \left(- (1+\lambda r^2) \frac{d^2}{dr^2} - (d-1+d\lambda r^2) \frac{1}{r} \frac{d}{dr} + \frac{l(l+d-2)}{r^2} + V_0(r)
  \right) \psi(r) = E \psi(r),  \label{eq:SE-d}
\end{equation}
where $\hat{J}^2$ has been replaced by its eigenvalues $l(l+d-2)$, $l=0,1, 2, \ldots$. The variable $r$ runs over $(0,+\infty)$ or $\left(0, 1/\sqrt{|\lambda|}\right)$ according to whether $\lambda>0$ or $\lambda<0$ and the differential operator in (\ref{eq:SE-d}) is formally self-adjoint with respect to the measure $d\mu = (1+\lambda r^2)^{-1/2} r^{d-1}dr$.\par
%
%
Equation (\ref{eq:SE-d}) is exactly solvable and its bound-state solutions can be expressed in terms of Jacobi polynomials as \cite{cq16}
\begin{equation}
  \psi_{n_r,l}(r) \propto r^l (1+\lambda r^2)^{-\beta/(2\lambda)} P_{n_r}^{\left(l+\frac{d-2}{2}, -\frac{\beta}{\lambda} -
  \frac{1}{2}\right)}(1+2\lambda r^2), \qquad n_r=0, 1, 2, \ldots,  \label{eq:wf-d}
\end{equation}
with corresponding energy eigenvalues
\begin{equation}
  E_n = \beta(2n+d) - \lambda n(n+d-1), \qquad n=2n_r+l.
\end{equation}
Here the range of $n$ values is determined by the normalizability of the radial wavefunctions $\psi_{n_r,l}(r)$ on the appropriate interval with respect to the measure $d\mu$ and is given by
\begin{equation}
  n =
    \begin{cases}
      0, 1, 2, \ldots, n_{\rm max}, \quad \frac{\beta}{\lambda}-\frac{d+1}{2} \le n_{\rm max} < \frac{\beta}{\lambda}
        - \frac{d-1}{2}, & \text{if $\lambda>0$,} \\[0.2cm]
      0, 1, 2, \ldots & \text{if $\lambda<0$.}
    \end{cases}  \label{eq:n-d}
\end{equation}
\par
%
%
At this stage, it is worth observing that the potentials $V_0(x)$ and $V_0(r)$ may be rewritten as
\begin{equation}
  V_0(x) = \lambda A - \frac{\lambda A}{1+\lambda x^2}, \qquad V_0(r) = \lambda A - \frac{\lambda A}{1+\lambda r^2},
  \label{eq:V_0}
\end{equation}
where, in both cases, $A = \frac{\beta}{\lambda}\left(\frac{\beta}{\lambda}+1\right)$. It is the form (\ref{eq:V_0}) that we will adopt in Sections III and IV.\par 
%
%
{}Furthermore, we may retrieve the one-dimensional results, contained in Eqs.~(\ref{eq:SE-1})--(\ref{eq:n-1}), from the $d$-dimensional ones, given in Eqs.~(\ref{eq:SE-d})--(\ref{eq:n-d}), by performing the changes
\begin{equation}
  d \rightarrow 1, \qquad r \rightarrow x, \qquad l \rightarrow p=0,1,  \label{eq:d->1}
\end{equation}
with $p$ related to the parity $(-1)^p = +1, -1$, and by extending the range of the variable from $(0, +\infty)$ or $\left(0, 1/\sqrt{|\lambda|}\right)$ to $(-\infty,+\infty)$ or $\left(-1/\sqrt{|\lambda|},1/\sqrt{|\lambda|}\right)$ according to the sign of $\lambda$. From Eqs.~(22.5.22) and (22.5.21) of Ref.~\cite{abramowitz}, we indeed note that after such substitutions, $P_{n_r}^{\left(l+\frac{d-2}{2}, -\frac{\beta}{\lambda} -\frac{1}{2}\right)}(1+2\lambda r^2)$ becomes
\begin{equation}
  P_{n/2}^{\left(-\frac{1}{2}, -\frac{\beta}{\lambda} -\frac{1}{2}\right)}(1+2\lambda r^2) \propto
  C_n^{(-\beta/\lambda)}\left(\sqrt{-\lambda r^2}\right) \qquad \text{for $p=0$, $n=2n_r$},
\end{equation}
or
\begin{equation}
  P_{(n-1)/2}^{\left(\frac{1}{2}, -\frac{\beta}{\lambda} -\frac{1}{2}\right)}(1+2\lambda r^2) \propto
  \frac{C_n^{(-\beta/\lambda)}\left(\sqrt{-\lambda r^2}\right)}{\sqrt{-\lambda r^2}} \qquad \text{for $p=1$, 
  $n=2n_r+1$},
\end{equation}
so that Eq.~(\ref{eq:wf-1}) directly follows from Eq.~(\ref{eq:wf-d}).\par
%
%
Since a similar replacement is valid for the extended potentials to be considered in Sections III and IV, we will only write there the $d$-dimensional results explicitly.\par
%
%
\section{FIRST FAMILY OF QES POTENTIALS}
\setcounter{equation}{0}

In Eq.~(\ref{eq:SE-d}), let us replace the potential $V_0(r)$ by a potential of the family
\begin{equation}
  V^{(1)}_m(r) = \lambda A - \frac{\lambda A}{1+\lambda r^2} + \lambda \sum_{k=1}^{2m} B_k (1+\lambda r^2)^k,
  \qquad m=1, 2, 3, \ldots,
\end{equation}
where $A, B_1, B_2, \ldots, B_{2m}$ are $2m+1$ parameters and the range of the variable $r$ is the same as in Section~II. In the resulting equation, let us make the changes of variable and of function
\begin{equation}
  \psi(r) = r^l \tilde{\psi}(z), \qquad z = \frac{1}{1+\lambda r^2},  \label{eq:var-function}
\end{equation}
where $0<z<1$ if $\lambda>0$ or $1<z<\infty$ if $\lambda<0$. This yields the following differential equation for $\tilde{\psi}(z)$,
\begin{equation}
  \left(-4z^2(1-z) \frac{d^2}{dz^2} + 2z(3z+2l+d-3) \frac{d}{dz} - Az + \sum_{k=1}^{2m} \frac{B_k}{z^k} - 
  \epsilon \right)\tilde{\psi}(z) = 0,  \label{eq:eq-V^{(1)}}
\end{equation}
where we have set 
\begin{equation}
  E = \lambda [\epsilon + A - l(l+d-1)].  \label{eq:E-epsilon}
\end{equation}
\par
%
%
After a brief inspection of Eq.~(\ref{eq:eq-V^{(1)}}), we make the transformation
\begin{equation}
  \tilde{\psi}(z) = z^a \exp\left(- \sum_{j=1}^m \frac{b_j}{z^j}\right) \phi(z),
\end{equation}
in terms of $m+1$ parameters $a, b_1, b_2, \ldots, b_m$, related to the previous ones. The resulting equation
\begin{align}
  & \biggl\{-4z^2(1-z) \frac{d^2}{dz^2} + 2 \biggl[(4a+3)z^2 - (4a-4b_1-2l-d+3)z - 4 \sum_{j=1}^m \frac{jb_j}{z^{j-1}}
     \nonumber \\
  & + 4 \sum_{j=2}^m \frac{jb_j}{z^{j-2}}\biggr] \frac{d}{dz}+ [2a(2a+1) - A]z - 4a^2 + 8ab_1 + 2a(2l+d-1) - 2b_1 
     - \epsilon \nonumber \\
  & - 4 \sum_{j=1}^m \frac{j(2a-j-1)b_j}{z^j} + 4 \sum_{j=2}^m \frac{j(2a-j-1)b_j}{z^{j-1}} - 4 \sum_{j=1}^m
     \frac{j^2b_j^2}{z^{2j}} + 4 \sum_{j=1}^m \frac{j^2b_j^2}{z^{2j-1}} \nonumber \\
  & - 8 \sum_{\substack{
        i, j=1\\
        i<j}}^m \frac{ijb_ib_j}{z^{i+j}} + 8 \sum_{\substack{
        i, j=1\\
        i<j}}^m \frac{ijb_ib_j}{z^{i+j-1}} + 6 \sum_{j=2}^m \frac{jb_j}{z^{j-1}} + 2(2l+d-3) \sum_{j=1}^m \frac{jb_j}{z^j}
        \nonumber \\
   & + \sum_{k=1}^{2m} \frac{B_k}{z^k} \biggr\} \phi(z) = 0  \label{eq:eq-V^{(1)}-bis}
\end{align}
looks rather complicated, but can be drastically simplified when considering the first few $m$ values, as we will now proceed to show.\par
%
%
\subsection{First potential of the first family}

{}For the first potential of the family
\begin{equation}
  V^{(1)}_1(r) = \lambda A - \frac{\lambda A}{1+\lambda r^2} + \lambda B_1 (1+\lambda r^2) + \lambda B_2
  (1+\lambda r^2)^2
\end{equation}
and the gauge transformation
\begin{equation}
  \tilde{\psi}(z) = z^a \exp\left(- \frac{b_1}{z}\right) \phi(z),
\end{equation}
Eq.~(\ref{eq:eq-V^{(1)}-bis}) reduces to
\begin{align}
  & \biggl\{-4z^2(1-z) \frac{d^2}{dz^2} + 2[(4a+3)z^2 - (4a-4b_1-2l-d+3)z - 4b_1] \frac{d}{dz} \nonumber \\
  & + [2a(2a+1) - A]z - 4a^2 + 8ab_1 + 2a(2l+d-1) - 2b_1 - \epsilon \nonumber \\
  & + \frac{-8(a-1)b_1 + 4b_1^2 + 2b_1(2l+d-3) + B_1}{z} + \frac{-4b_1^2 + B_2}{z^2}\biggr\} \phi(z) = 0.
\end{align}
With the constraints
\begin{equation}
  B_1 = 2b_1(4a-2b_1-2l-d-1), \qquad B_2 = 4b_1^2,  \label{eq:C-V^{(1)}_1}
\end{equation}
determining $B_1$ and $B_2$ in terms of $a$ and $b_1$ (or vice versa), we obtain the equation
\begin{align}
  & \biggl\{-4z^2(1-z) \frac{d^2}{dz^2} + 2[(4a+3)z^2 - (4a-4b_1-2l-d+3)z - 4b_1] \frac{d}{dz} \nonumber \\
  & + [2a(2a+1) - A]z - 4a^2 + 8ab_1 + 2a(2l+d-1) - 2b_1 - \epsilon \biggr\} \phi(z) = 0.  \label{eq:eq-V^{(1)}-ter}
\end{align}
\par
%
%
Provided some additional constraints are satisfied, the latter has exact solutions that are $n$th-degree polynomials in $z$,
\begin{equation}
  \phi_n(z) = \prod_{i=1}^n (z-z_i), \qquad n\ne 0, \qquad \phi_0(z) = 1,  \label{eq:polynomial}
\end{equation}
with distinct roots $z_i$, $i=1, 2, \ldots, n$. Substituting (\ref{eq:polynomial}) into (\ref{eq:eq-V^{(1)}-ter}) and applying the functional Bethe ansatz method \cite{zhang, agboola13, agboola14}, outlined in the Appendix, we obtain
\begin{equation}
\begin{split}
  A & = (2a+2n)(2a+2n+1), \\
  E_{n,l} & = \lambda\Bigl[2(4a+4n-1) \sum_i z_i + 8(a+n)b_1 + 2(a+n)(2l+d) - 2b_1 \\ 
  & \quad {}- l(l+d-1)\Bigr],
\end{split}
\end{equation}
where the roots $z_i$ satisfy the Bethe ansatz equations
\begin{equation}
  \sum_{j\ne i} \frac{2}{z_i-z_j} + \frac{(4a+3)z_i^2 - (4a-4b_1-2l-d+3)z_i - 4b_1}{2z_i^2(z_i-1)} = 0, \qquad i=1, 2, 
  \ldots, n.
\end{equation}
The corresponding wavefunctions 
\begin{equation}
  \psi_{n,l}(r) \propto r^l (1+\lambda r^2)^{-a} \exp[-b_1(1+\lambda r^2)] \phi_n\left(\frac{1}{1+\lambda r^2}\right)
\end{equation}
are normalizable with respect to the measure $d\mu = (1+\lambda r^2)^{-1/2} r^{d-1} dr$ provided $b_1>0$ if $\lambda>0$ or $a<\frac{1}{4}-n$ if $\lambda<0$.\par
%
%
As examples of the above general expressions of the exact solutions, let us consider the first two $n$ values. For $n=0$, we get
\begin{equation}
\begin{split}
  & A = 2a(2a+1), \qquad E_{0,l} = \lambda[8ab_1 + 2a(2l+d) - 2b_1 - l(l+d-1)], \\
  & \psi_{0,l}(r) \propto r^l (1+\lambda r^2)^{-a} \exp(-\lambda b_1 r^2),
\end{split}  \label{eq:results}
\end{equation}
and for $n=1$, 
\begin{equation}
\begin{split}
  & A = (2a+2)(2a+3), \\
  & E_{1,l} = \lambda[2(4a+3)z_1 + 8(a+1)b_1 + 2(a+1)(2l+d) - 2b_1 - l(l+d-1)], \\
  & \psi_{1,l}(r) \propto r^l (1+\lambda r^2)^{-a-1} [1 - z_1(1+\lambda r^2)] \exp(-\lambda b_1 r^2),
\end{split}
\end{equation}
where the root $z_1$ is determined by the Bethe ansatz equation
\begin{equation}
  (4a+3)z_1^2 - (4a-4b_1-2l-d+3)z_1 - 4b_1 = 0,
\end{equation}
yielding
\begin{equation}
\begin{split}
  & z_1 = \frac{1}{4a+3} \left(2a - 2b_1 - l - \frac{d-3}{2} \pm \Delta\right), \\
  & \Delta = \left[4(a+b_1)^2 + 4(a+2b_1) + 1 - \left(l + \frac{d-1}{2}\right)\left(4a - 4b_1 - l - \frac{d-5}{2}\right)
     \right]^{1/2},
\end{split}  \label{eq:results-bis}
\end{equation}
provided $\Delta$ is real. The function $\psi_{0,l}(r)$ having no node for the allowed values of $r$ represents a ground state, while $\psi_{1,l}(r)$ may describe a ground or first-excited state according to the value taken by $z_1$.\par
%
%
It is worth observing that Eq.~(\ref{eq:eq-V^{(1)}-ter}) being of Heun type, the differential equation corresponding to $V^{(1)}_1(r)$ with the constraints (\ref{eq:C-V^{(1)}_1}) and $A = (2a+2n)(2a+2n+1)$ has a hidden sl(2,$\R$) symmetry. It can indeed be rewritten in terms of an element of the sl(2,$\R$) enveloping algebra acting on $\phi_n(z)$,
\begin{align}
  & \Bigl[4 J_n^+ J_n^0 - 4 J_n^+ J_n^- + 2(4a+1+3n) J_n^+ - 2(4a-4b_1-2l-d+3+2n) J_n^0 \nonumber \\
  & - 8b_1 J_n^- - 4a^2 + 8ab_1 + 2a(2l+d-1) - 2b_1 - n(4a-4b_1-2l-d+3+2n)\Bigr] \phi_n(z) \nonumber \\
  & = \epsilon_n \phi_n(z),  \label{eq:sl-1}
\end{align}
where
\begin{equation}
  J_n^+ = z^2 \frac{d}{dz} - nz, \qquad J_n^0 = z \frac{d}{dz} - \frac{n}{2}, \qquad J_n^- = \frac{d}{dz}
  \label{eq:sl(2,R)}
\end{equation}
are differential operator realizations of the sl(2,$\R$) generators in the $(n+1)$-dimensional representation spanned by ${\cal P}_{n+1} = (1, z, z^2, \ldots, z^n)$. On taking into account that $J_n^+ z^k = (k-n) z^{k+1}$, $J_n^0 z^k = \left(k - \frac{n}{2}\right) z^k$, $J_n^- z^k = k z^{k-1}$, it is straightforward to write the matrix of the differential operator on the left-hand side of (\ref{eq:sl-1}) explicitly. Its diagonalization then provides the admissible energies $\epsilon_n$ and corresponding eigenfunctions $\phi_n(z)$. In this way it is possible to rederive the results contained in Eqs.~(\ref{eq:results})--(\ref{eq:results-bis}). In the terminology used for QES problems with a hidden sl(2,$\R$) symmetry \cite{turbiner16}, the Schr\"odinger equation corresponding to the potential $V^{(1)}_1(r)$ is a QES equation of the first type, the parameter $\epsilon$ in Eq.~(\ref{eq:eq-V^{(1)}}) playing the role of energy $E$ (see Eq.~(\ref{eq:E-epsilon})).\par
%
%
\subsection{Second potential of the first family}

{}For the second potential of the family
\begin{equation}
  V^{(1)}_2(r) = \lambda A - \frac{\lambda A}{1+\lambda r^2} + \lambda B_1 (1+\lambda r^2) + \lambda B_2
  (1+\lambda r^2)^2 + \lambda B_3 (1+\lambda r^2)^3 + \lambda B_4 (1+\lambda r^2)^4
\end{equation}
and the gauge transformation
\begin{equation}
  \tilde{\psi}(z) = z^a \exp\left(- \frac{b_1}{z} - \frac{b_2}{z^2}\right) \phi(z),
\end{equation}
we get the equation
\begin{align}
  & \biggl\{-4z^3(1-z) \frac{d^2}{dz^2} + 2[(4a+3)z^3 - (4a-4b_1-2l-d+3)z^2 - 4(b_1-2b_2)z - 8b_2] \frac{d}{dz}
     \nonumber \\
  & + [2a(2a+1) - A]z^2 + [- 4a^2 + 8ab_1 + 2a(2l+d-1) - 2b_1 - \epsilon]z + B_1 \nonumber \\
  & - 2b_1(4a-2b_1-2l-d-1) + 4b_2(4a-3) \biggr\} \phi(z) = 0
\end{align} 
after imposing the constraints
\begin{equation}
  B_2 = 4[b_1^2 + b_2(4a-4b_1-2l-d-3)], \qquad B_3 = 16b_2(b_1-b_2), \qquad B_4 = 16b_2^2, 
\end{equation}
expressing $B_2$, $B_3$, and $B_4$ in terms of $a$, $b_1$, and $b_2$ (or vice versa).\par
%
%
This time, we obtain the relations
\begin{equation}
\begin{split}
  A &= (2a+2n)(2a+2n+1), \\
  B_1 &= - 2(4a+4n-1) \sum_i z_i^2 - 8 \sum_{i<j} z_i z_j + 2(4a+4n-4b_1-2l-d-1) \sum_i z_i \\
  &\quad {}+ 2b_1(4a+4n-2b_1-2l-d-1) - 4b_2(4a+4n-3), \\
  E_{n,l} &= \lambda [2(4a+4n-1) \sum_i z_i + 8(a+n)b_1 + 2(a+n)(2l+d) - 2b_1\\ 
  &\quad {}- l(l+d-1)],
\end{split}
\end{equation}
and the Bethe ansatz equations
\begin{align}
  &\sum_{j\ne i} \frac{2}{z_i-z_j} + \frac{(4a+3)z_i^3 - (4a-4b_1-2l-d+3)z_i^2 - 4(b_1-2b_2)z_i - 8b_2}{2z_i^3(z_i-1)} 
  = 0, \nonumber \\ 
  & \quad i=1, 2, \ldots, n.
\end{align}
The normalization condition for the wavefunctions $\psi_{n,l}(r)$ is now $b_2>0$ if $\lambda>0$ or $a<\frac{1}{4}-n$ if $\lambda<0$.\par
%
%
As before, for $n=0$, we get the ground-state energy and wavefunction
\begin{equation}
\begin{split}
  & E_{0,l} = \lambda[8ab_1 + 2a(2l+d) - 2b_1 - l(l+d-1)], \\
  & \psi_{0,l}(r) \propto r^l (1+\lambda r^2)^{-a} \exp[-\lambda (b_1+2b_2) r^2 - \lambda^2 b_2 r^4],
\end{split} 
\end{equation}
with the constraints
\begin{equation}
  A = 2a(2a+1), \qquad B_1 = 2b_1(4a-2b_1-2l-d-1) - 4b_2(4a-3).
\end{equation}
{}For $n=1$,
\begin{equation}
\begin{split}
  A &= (2a+2)(2a+3), \\
  B_1 &= - 2(4a+3)z_1^2 + 2(4a-4b_1-2l-d+3)z_1 + 2b_1(4a-2b_1-2l-d+3) \\
  & \quad {} - 4b_2(4a+1),
\end{split}
\end{equation}
and $z_1$ a real solution of
\begin{equation}
  (4a+3)z_1^3 - (4a-4b_1-2l-d+3)z_1^2 - 4(b_1-2b_2)z_1 - 8b_2 = 0,
\end{equation}
hence, for instance,
\begin{equation}
\begin{split}
  z_1 &= \frac{4a-4b_1-2l-d+3}{3(4a+3)} + \left(- \frac{v}{2} + \sqrt{\left(\frac{v}{2}\right)^2 + 
    \left(\frac{u}{3}\right)^3}\right)^{1/3} \\
  &\quad {} + \left(- \frac{v}{2} - \sqrt{\left(\frac{v}{2}\right)^2 + \left(\frac{u}{3}\right)^3}\right)^{1/3}, \\
  u &= \frac{4}{4a+3} \left(- b_1 + 2b_2 - \frac{(4a-4b_1-2l-d+3)^2}{12(4a+3)}\right), \\
  v &= - \frac{8}{4a+3} \biggl(b_2 + \frac{(4a-4b_1-2l-d+3)^3}{108(4a+3)^2} \\
  &\quad {} + \frac{(4a-4b_1-2l-d+3)(b_1-2b_2)}{6(4a+3)}\biggr),
\end{split}
\end{equation}
we obtain
\begin{equation}
\begin{split}
  & E_{1,l} = \lambda[2(4a+3)z_1 + 8(a+1)b_1 + 2(a+1)(2l+d) - 2b_1 - l(l+d-1)], \\
  & \psi_{1,l}(r) \propto r^l (1+\lambda r^2)^{-a-1} [1-z_1(1+\lambda r^2)] \exp{[- \lambda(b_1+2b_2)r^2 
    - \lambda^2 b_2 r^4]},
\end{split}
\end{equation}
which may correspond to a ground or first-excited state according to the value of $z_1$.\par
%
%
\subsection{Third potential of the first family}

{}For 
\begin{equation}
\begin{split}
  V^{(1)}_3(r) &= \lambda A - \frac{\lambda A}{1+\lambda r^2} + \lambda B_1 (1+\lambda r^2) + \lambda B_2
    (1+\lambda r^2)^2 + \lambda B_3 (1+\lambda r^2)^3 \\  
  &\quad {} + \lambda B_4 (1+\lambda r^2)^4 + \lambda B_5 (1+\lambda r^2)^5 + \lambda B_6 (1+\lambda r^2)^6
\end{split}
\end{equation}
and the gauge transformation
\begin{equation}
  \tilde{\psi}(z) = z^a \exp\left(- \frac{b_1}{z} - \frac{b_2}{z^2} - \frac{b_3}{z^3}\right) \phi(z),
\end{equation}
the constraints
\begin{equation}
\begin{split}
  B_3 &= 2[8b_2(b_1-b_2) + 3b_3(4a-4b_1-2l-d-5)], \qquad B_4 = 8[2b_2^2 + 3b_3(b_1-2b_2)], \\
  B_5 &= 12b_3(4b_2-3b_3), \qquad B_6 = 36b_3^2,
\end{split}
\end{equation}
expressing $B_3$, $B_4$, $B_5$, and $B_6$ in terms of $a$, $b_1$, $b_2$, and $b_3$ (or vice versa), lead to the equation
\begin{equation}
\begin{split}
  & \biggl\{- 4z^4(1-z) \frac{d^2}{dz^2} + 2[(4a+3)z^4 - (4a-4b_1-2l-d+3)z^3 - 4(b_1-2b_2)z^2 \\
  & - 4(2b_2-3b_3)z -12b_3] \frac{d}{dz} + [2a(2a+1) - A]z^3 \\
  & + [- 4a^2 + 8ab_1 + 2a(2l+d-1) - 2b_1 - \epsilon]z^2 + [B_1 - 2b_1(4a-2b_1-2l-d-1) \\
  & + 4b_2(4a-3)]z + B_2 - 4b_1^2 - 4b_2(4a-4b_1-2l-d-3) + 6b_3(4a-5)\biggr\} \phi(z) = 0.
\end{split}
\end{equation}
\par
%
%
We now impose the relations
\begin{equation}
\begin{split}
  A &= (2a+2n)(2a+2n+1), \\
  B_1 &= - 2(4a+4n-1) \sum_i z_i^2 - 8 \sum_{i<j} z_i z_j + 2(4a+4n-4b_1-2l-d-1) \sum_i z_i \\
  &\quad {} + 2b_1(4a+4n-2b_1-2l-d-1) - 4b_2(4a+4n-3), \\
  B_2 &= -2(4a+4n-1) \sum_i z_i^3 - 8 \sum_{i\ne j} z_i^2 z_j + 2(4a+4n-4b_1-2l-d-1) \sum_i z_i^2 \\
  &\quad {}+ 8 \sum_{i<j} z_i z_j + 8(b_1-2b_2) \sum_i z_i + 4b_1^2 + 4b_2(4a+4n-4b_1-2l-d-3) \\
  &\quad {}- 6b_3(4a+4n-5), \\
  E_{n,l} &= \lambda[2(4a+4n-1) \sum_i z_i + 8(a+n)b_1 + 2(a+n)(2l+d) \\
  &\quad {} - 2b_1 - l(l+d-1)], 
\end{split}
\end{equation}
and the Bethe ansatz equations
\begin{align}
  &\sum_{j\ne i} \frac{2}{z_i-z_j} \nonumber \\
  &{} + \frac{(4a+3)z_i^4 - (4a-4b_1-2l-d+3)z_i^3 - 4(b_1-2b_2)z_i^2 - 4(2b_2-3b_3)z_i
    - 12b_3}{2z_i^4(z_i-1)} = 0, \nonumber \\ 
  & \quad i=1, 2, \ldots, n,
\end{align}
as well as the normalization condition $b_3>0$ if $\lambda>0$ or $a<\frac{1}{4}-n$ if $\lambda<0$.\par
%
%
{}For $n=0$, we get the ground-state energy and wavefunction
\begin{equation}
\begin{split}
  & E_{0,l} = \lambda[8ab_1 + 2a(2l+d) - 2b_1 - l(l+d-1)], \\
  & \psi_{0,l}(r) \propto r^l (1+\lambda r^2)^{-a} \exp[-\lambda (b_1+2b_2+3b_3) r^2 - \lambda^2(b_2+3b_3) r^4
    - \lambda^3 b_3 r^6],
\end{split} 
\end{equation}
together with the constraints
\begin{equation}
\begin{split}
  A &= 2a(2a+1), \\
  B_1 &= 2b_1(4a-2b_1-2l-d-1) - 4b_2(4a-3), \\
  B_2 &= 4b_1^2 + 4b_2(4a-4b_1-2l-d-3) - 6b_3(4a-5).
\end{split}
\end{equation}
For $n=1$, we obtain
\begin{equation}
\begin{split}
  E_{1,l} &= \lambda[2(4a+3)z_1 + 8(a+1)b_1 + 2(a+1)(2l+d) - 2b_1 - l(l+d-1)], \\
  \psi_{1,l}(r) &\propto r^l (1+\lambda r^2)^{-a-1} [1-z_1(1+\lambda r^2)] \\
  &\quad \times \exp{[- \lambda(b_1+2b_2+3b_3)r^2 - \lambda^2 (b_2+3b_3) r^4 - \lambda^3 b_3 r^6]},
\end{split}
\end{equation}
which may correspond to a ground or first-excited state again. Here, the constraints read
\begin{equation}
\begin{split}
  A &= (2a+2)(2a+3), \\
  B_1 &= - 2(4a+3)z_1^2 + 2(4a-4b_1-2l-d+3)z_1 + 2b_1(4a-2b_1-2l-d+3) \\
  &\quad {} - 4b_2(4a+1), \\
  B_2 &= - 2(4a+3)z_1^3 + 2(4a-4b_1-2l-d+3)z_1^2 + 8(b_1-2b_2)z_1 + 4b_1^2 \\
  &\quad {} + 4b_2(4a-4b_1-2l-d+1) - 6b_3(4a-1),
\end{split}
\end{equation}
where $z_1$ is a real solution of the equation
\begin{equation}
  (4a+3)z_1^4 - (4a-4b_1-2l-d+3)z_1^3 - 4(b_1-2b_2)z_1^2 - 4(2b_2-3b_3)z_1 - 12b_3 = 0.
\end{equation}
\par
%
%
\subsection{The one-dimensional case}

As mentioned in Section II for the oscillator alone, the one-dimensional case on the real line for the extended oscillator can be easily retrieved from the $d$-dimensional one on the half-line for $d\ge 2$ by applying (\ref{eq:d->1}) and extending the range of the variable accordingly. In the transformed wavefunctions, $p=0,1$ is related to the parity $(-1)^p = +1, -1$. The normalization conditions with respect to the measure $d\mu = (1+\lambda x^2)^{-1/2} dx$ remain the same as in $d\ge 2$ dimensions with respect to the measure $d\mu = (1+\lambda r^2)^{-1/2} r^{d-1} dr$.\par
%
%
{}For instance, for the first potential $V^{(1)}_1(x)$, with the constraints
\begin{equation}
  A = 2a(2a+1), \qquad B_1 = 4b_1(2a-b_1-1-p), \qquad B_2 = 4b_1^2,
\end{equation}
we obtain from (\ref{eq:results})
\begin{equation}
\begin{split}
  &E_{0,p} = \lambda(8ab_1 + 4pa + 2a - 2b_1 - p), \\
  &\psi_{0,p}(x) \propto x^p (1+\lambda x^2)^{-a} \exp(-\lambda b_1 x^2),
\end{split}
\end{equation}
where $b_1>0$ if $\lambda>0$ or $a<\frac{1}{4}$ if $\lambda<0$. These energy and wavefunction correspond to a ground state for $p=0$ or to a first-excited state for $p=1$.\par
%
%
In Fig.~1, some examples of extended potentials are plotted and compared with $V_0(x)$ for $\lambda>0$. The corresponding wavefunctions are displayed in Fig.~2. Figures 3 and 4 show similar comparisons in the $\lambda<0$ case.\par
%
%
\begin{figure}[h]
\begin{center}
\includegraphics{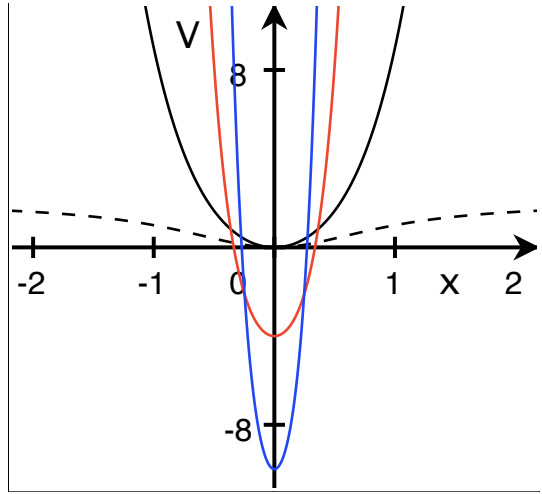}
\caption{Plot of $V^{(1)}_m(x)$ as a function of $x$ for $m=1$, $b_1=1$ (black solid line), $m=2$, $b_1=b_2=1$ (red line), and $m=3$, $b_1=b_2=b_3=1$ (blue line). In all cases, $\lambda=1$, $a=1/2$, $n=p=0$ (hence $A=2$), and $E_{0,0}=3$. The potential $V_0(x)$ for $\lambda=1$, $A=2$, and ground-state energy $E_0=1$ is also shown (black dashed line).}
\end{center}
\end{figure}
\par
%
%
\begin{figure}[h]
\begin{center}
\includegraphics{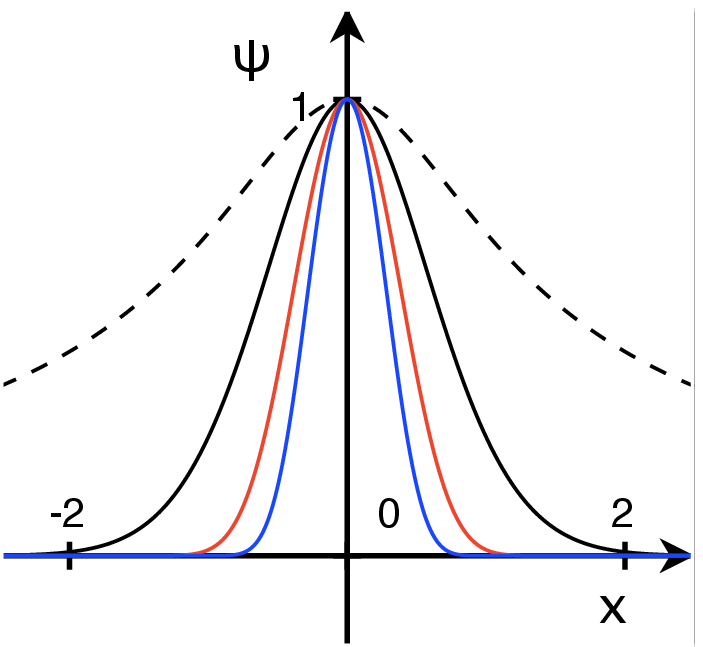}
\caption{Plot of ground-state wavefunction $\psi_{0,0}(x)$ of $V^{(1)}_m(x)$ as a function of $x$ for $m=1$, $b_1=1$ (black solid line), $m=2$, $b_1=b_2=1$ (red line), and $m=3$, $b_1=b_2=b_3=1$ (blue line). In all cases, $\lambda=1$, $a=1/2$, $n=p=0$ (hence $A=2$), and $E_{0,0}=3$. The ground-state wavefunction $\psi_0(x)$ of $V_0(x)$ for $\lambda=1$, $A=2$, and $E_0=1$ is also shown (black dashed line).}
\end{center}
\end{figure}
\par
%
%
\begin{figure}[h]
\begin{center}
\includegraphics{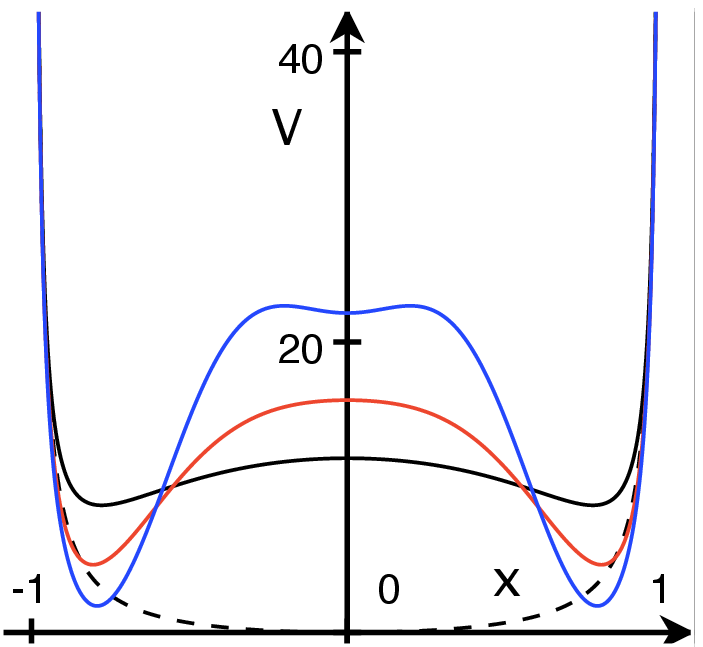}
\caption{Plot of $V^{(1)}_m(x)$ as a function of $x$ for $m=1$, $b_1=1$ (black solid line), $m=2$, $b_1=b_2=1$ (red line), and $m=3$, $b_1=b_2=b_3=1$ (blue line). In all cases, $\lambda=-1$, $a=-1$, $n=p=0$ (hence $A=2$), and $E_{0,0}=12$. The potential $V_0(x)$ for $\lambda=-1$, $A=2$, and ground-state energy $E_0=2$ is also shown (black dashed line).}
\end{center}
\end{figure}
\par
%
%
\begin{figure}[h]
\begin{center}
\includegraphics{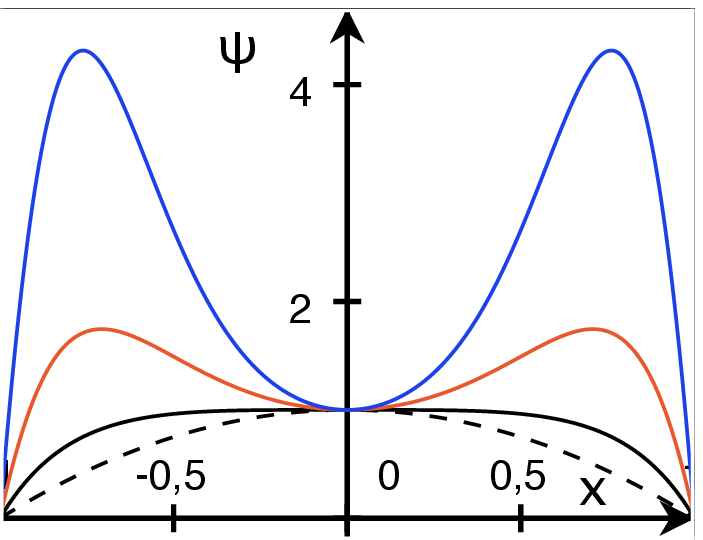}
\caption{Plot of ground-state wavefunction $\psi_{0,0}(x)$ of $V^{(1)}_m(x)$ as a function of $x$ for $m=1$, $b_1=1$ (black solid line), $m=2$, $b_1=b_2=1$ (red line), and $m=3$, $b_1=b_2=b_3=1$ (blue line). In all cases, $\lambda=-1$, $a=-1$, $n=p=0$ (hence $A=2$), and $E_{0,0}=12$. The ground-state wavefunction $\psi_0(x)$ of $V_0(x)$ for $\lambda=-1$, $A=2$, and $E_0=2$ is also shown (black dashed line).}
\end{center}
\end{figure}
\par
%
%
%
%
\section{SECOND FAMILY OF QES POTENTIALS}
\setcounter{equation}{0}

Instead of $V_0(r)$ in Eq.~(\ref{eq:SE-d}), let us now consider a potential of the family
\begin{equation}
  V^{(2)}_m(r) = \lambda A - \frac{\lambda A}{1+\lambda r^2} + \lambda \sum_{k=1}^{2m} \frac{B_k}
  {(1+\lambda r^2)^{k+1}},  \qquad m=1, 2, 3, \ldots,
\end{equation}
where $A, B_1, B_2, \ldots, B_{2m}$ are $2m+1$ parameters and the range of the variable $r$ is the same as in Section~II. The transformations in Eq.~(\ref{eq:var-function}) now yield
\begin{equation}
  \left(-4z^2(1-z) \frac{d^2}{dz^2} + 2z(3z+2l+d-3) \frac{d}{dz} - Az + \sum_{k=1}^{2m} B_k z^{k+1} - 
  \epsilon \right)\tilde{\psi}(z) = 0,  \label{eq:eq-V^{(2)}} 
\end{equation}
where $\epsilon$ is defined in terms of $E$ as in Eq.~(\ref{eq:E-epsilon}). After a brief inspection of Eq.~(\ref{eq:eq-V^{(2)}}), we make the substitution
\begin{equation}
  \tilde{\psi}(z) = z^a \exp\left(- \sum_{j=1}^m b_j z^j\right) \phi(z),
\end{equation}
where $a, b_1, b_2, \ldots, b_m$ are $m+1$ parameters related to the previous ones. Setting 
\begin{equation}
  \epsilon = - 2a(2a-2l-d+1)  \label{eq:epsilon}
\end{equation}
at the same time also leads to a drastic simplification. The resulting differential equation for $\phi(z)$ reads
\begin{align}
  & \biggl\{-4z(1-z) \frac{d^2}{dz^2} + 2 \biggl[- 4a+2l+d-3 + (4a+4b_1+3)z + 4 \sum_{j=2}^m jb_j z^j \nonumber \\
  & - 4 \sum_{j=1}^m jb_j z^{j+1}\biggr] \frac{d}{dz} + 2a(2a+1) + 2b_1(4a-2l-d+3) - A \nonumber \\
  & + [-2b_1(4a+4b_1+3) + B_1]z + 4 \sum_{j=2}^m j(2a+j-1)b_j z^{j-1} - 4 \sum_{j=2}^m j(2a+j-1)b_j z^j 
    \nonumber \\
  & - 4 \sum_{j=2}^m j^2 b_j^2 z^{2j-1} + 4 \sum_{j=1}^m j^2 b_j^2 z^{2j} - 8 \sum_{\substack{
        i, j=1\\
        i<j}}^m ijb_ib_j z^{i+j-1} + 8 \sum_{\substack{
        i, j=1\\
        i<j}}^m ijb_ib_j z^{i+j} \nonumber \\
  & - 6 \sum_{j=2}^m jb_j z^j - 2(2l+d-3) \sum_{j=2}^m jb_j z^{j-1} + \sum_{k=2}^{2m} B_k z^k\biggr\} \phi(z) = 0.
\end{align} 
As in Section III, we will now consider successively the $m=1$, 2, and 3 cases.\par
%
%
\subsection{First potential of the second family}

{}For the first potential of the family
\begin{equation}
  V^{(2)}_1(r) = \lambda A - \frac{\lambda A}{1+\lambda r^2} + \frac{\lambda B_1}{(1+\lambda r^2)^2} +
  \frac{\lambda B_2}{(1+\lambda r^2)^3},
\end{equation}
the gauge transformation
\begin{equation}
  \tilde{\psi}(z) = z^a \exp(-b_1z) \phi(z),
\end{equation}
and constraint (\ref{eq:epsilon}), the differential equation for $\phi(z)$ reads
\begin{align}
  & \biggl\{ - 4z(1-z) \frac{d^2}{dz^2} + 2 [- 4b_1z^2 + (4a+4b_1+3)z - 4a + 2l + d -3] \frac{d}{dz} \nonumber \\
  & + [B_1 - 2b_1(4a+2b_1+3)]z - A + 2a(2a+1) + 2b_1(4a-2l-d+3)\biggr\} \phi(z) = 0,  \label{eq:eq-V^(2)}
\end{align}
where we have also set
\begin{equation}
  B_2 = - 4b_1^2.  \label{eq:B_2}
\end{equation}
We note that the two constraints (\ref{eq:epsilon}) and (\ref{eq:B_2}) determine $\epsilon$ and $B_2$ in terms of $a$ and $b_1$ (or vice versa).\par
%
%
Equation (\ref{eq:eq-V^(2)}) has polynomial solutions of type (\ref{eq:polynomial}) provided the conditions
\begin{equation}
\begin{split}
  B_1 &= 2b_1(4a+4n+2b_1+3), \\
  A &= (2a+2n)(2a+2n+1) + 2b_1\Bigl(4a+4n-2l-d+3 - 4 \sum_i z_i\Bigr)
\end{split}  \label{eq:B_1-A}
\end{equation}
are fulfilled with $z_i$ satisfying the Bethe ansatz equations
\begin{equation}
  \sum_{j\ne i} \frac{2}{z_i-z_j} + \frac{- 4b_1z_i^2 + (4a+4b_1+3)z_i - 4a+2l+d-3}{2z_i(z_i-1)} = 0, \qquad
  i=1, 2, \ldots, n.
\end{equation}
On taking Eqs.~(\ref{eq:E-epsilon}), (\ref{eq:epsilon}), and (\ref{eq:B_1-A}) into account, the energy eigenvalues are given by
\begin{equation}
  E_{n,l} = \lambda \Bigl[2a(4n+2l+d) + 2n(2n+1) + 2b_1\Bigl(4a+4n-2l-d+3 - 4 \sum_i z_i\Bigr) - l(l+d-1)\Bigr].
\end{equation}
 The corresponding wavefunctions
 \begin{equation}
  \psi_{n,l}(r) \propto r^l (1+\lambda r^2)^{-a} \exp\left(- \frac{b_1}{1+\lambda r^2}\right) \phi_n\left(\frac{1}{1+\lambda
  r^2}\right)
\end{equation}
are normalizable with respect to the measure $d\mu = (1+\lambda r^2)^{-1/2} r^{d-1} dr$ provided $2a > l + \frac{d-1}{2}$ if $\lambda>0$ or $b_1>0$ if $\lambda<0$.\par
%
%
As examples, we find for $n=0$
\begin{equation}
\begin{split}
  &A = 2a(2a+1) +2b_1(4a-2l-d+3), \qquad B_1 = 2b_1(4a+2b_1+3), \\
  &E_{0,l} = \lambda [2a(2l+d) + 2b_1(4a-2l-d+3) - l(l+d-1)], \\
  &\psi_{0,l}(r) \propto r^l (1+\lambda r^2)^{-a} \exp\left(- \frac{b_1}{1+\lambda r^2}\right),
\end{split} \label{eq:results-ter}
\end{equation}
corresponding to a ground state, and for $n=1$
\begin{equation}
\begin{split}
  &A = (2a+2)(2a+3) +2b_1(4a-2l-d+7-4z_1), \qquad B_1 = 2b_1(4a+2b_1+7), \\
  &E_{1,l} = \lambda [2a(2l+d+4) + 2b_1(4a-2l-d+7-4z_1) - l(l+d-1) + 6], \\
  &\psi_{1,l}(r) \propto r^l (1+\lambda r^2)^{-a-1} [1 - z_1(1+\lambda r^2)] \exp\left(- \frac{b_1}{1+\lambda r^2}\right),
\end{split}
\end{equation}
corresponding to a ground or first-excited state. Here $z_1$ is determined by the Bethe ansatz equation
\begin{equation}
  4b_1z_1^2 - (4a+4b_1+3)z_1 + 4a-2l-d+3 = 0,
\end{equation}
hence
\begin{equation}
  z_1 = \frac{1}{8b_1} (4a+4b_1+3 \pm \Delta), \qquad \Delta = [16(a-b_1)^2 + 24a + 8(4l+2d-3)b_1 +9]^{1/2},
\end{equation}
provided $\Delta$ is real.\par
%
%
As in Section IIIA, the differential equation (\ref{eq:eq-V^(2)}) obtained here is of Heun type. Hence, on assuming $B_1 = 2b_1(4a+4n+2b_1+3)$, it is clear that it has a hidden sl(2,$\R$) symmetry. It can indeed be rewritten then as an element of the sl(2,$\R$) enveloping algebra acting on $\phi_n(z)$,
\begin{align}
  & \bigl[4J_n^+ J_n^- - 4 J_n^0 J_n^- - 8b_1 J_n^+ + 2(4a+4b_1+3+2n) J_n^0 - 2(4a-2l-d+3+n) J_n^- \nonumber \\
  & - A + 2a(2a+1) + 2b_1(4a-2l-d+3) + n(4a+4b_1+3+2n)\bigr] \phi_n(z) = 0,
\end{align}
with $J_n^+$, $J_n^0$, $J_n^-$ defined in (\ref{eq:sl(2,R)}). The diagonalization of such an operator in the space ${\cal P}_{n+1}$ provides the possible values of $A$, as given above. In the terminology used for QES problems with a hidden sl(2,$\R$) symmetry \cite{turbiner16}, the Schr\"odinger equation corresponding to the potential $V^{(2)}_1(r)$ is a QES equation of the second type: the parameter $\epsilon$, related to the energy $E$, is fixed through Eq.~(\ref{eq:epsilon}) and one of the potential parameters, namely $A$, plays the role of the eigenvalue. This is actually a generalization of the Sturm representation for the Coulomb problem, where the energy is fixed and the charge is quantized.\par
%
%
\subsection{Second potential of the second family}

Considering next the potential
\begin{equation}
  V^{(2)}_2(r) = \lambda A - \frac{\lambda A}{1+\lambda r^2} + \frac{\lambda B_1}{(1+\lambda r^2)^2} +
  \frac{\lambda B_2}{(1+\lambda r^2)^3} + \frac{\lambda B_3}{(1+\lambda r^2)^4} + \frac{\lambda B_4}
  {(1+\lambda r^2)^5}
\end{equation}
and the gauge transformation
\begin{equation}
  \tilde{\psi}(z) = z^a \exp(-b_1z - b_2z^2) \phi(z),
\end{equation}
we get the equation
\begin{align}
  & \biggl\{- 4z(1-z) \frac{d^2}{dz^2} + 2 [-8b_2z^3- 4(b_1-2b_2)z^2 + (4a+4b_1+3)z - 4a + 2l + d -3] \frac{d}{dz}
     \nonumber \\
  & + [B_2 + 4b_1^2 - 4b_2(4a+4b_1+5)]z^2 + [B_1 - 2b_1(4a+2b_1+3) + 4b_2(4a-2l-d+5)]z \nonumber \\
  & - A + 2a(2a+1) + 2b_1(4a-2l-d+3)\biggr\} \phi(z) = 0 
\end{align}
after imposing the constraints
\begin{equation}
  B_3 = - 16b_2(b_1-b_2), \qquad B_4 = - 16b_2^2,
\end{equation}
together with Eq.~(\ref{eq:epsilon}), thus expressing $\epsilon$, $B_3$, and $B_4$ in terms of $a$, $b_1$, and $b_2$ (or vice versa).\par
%
%
The Bethe ansatz method now leads to the relations
\begin{equation}
\begin{split}
  B_2 &= - 4b_1^2 + 4b_2(4a+4n+4b_1+5), \\
  B_1 &= 2b_1(4a+4n+2b_1+3) + 4b_2 \Bigl(4 \sum_i z_i - 4a-4n+2l+d-5\Bigr), \\
  A &= (2a+2n)(2a+2n+1) - 2b_1 \Bigl(4 \sum_i z_i - 4a-4n+2l+d-3\Bigr) \\
  &\quad {} - 16b_2 \Bigl(\sum_i z_i^2 - \sum_i z_i\Bigr),
\end{split}
\end{equation}
with the $z_i$'s solutions of
\begin{align}
  &\sum_{j\ne i} \frac{2}{z_i-z_j} + \frac{ - 8b_2z_i^3 - 4(b_1-2b_2)z_i^2 + (4a+4b_1+3)z_i - 4a+2l+d-3}{2z_i(z_i-1)} =
     0, \nonumber \\
  &\quad i=1, 2, \ldots, n.
\end{align}
Hence
\begin{align}
  E_{n,l} &= \lambda \Bigl[2a(4n+2l+d) + 2n(2n+1) + 2b_1\Bigl(4a+4n-2l-d+3-4\sum_i z_i\Bigr) \nonumber \\
  &\quad {}- 16b_2\Bigl(\sum_i z_i^2 - \sum_i z_i\Bigr) - l(l+d-1)\Bigr],
\end{align}
together with the normalization condition $2a>l+\frac{d-1}{2}$ if $\lambda>0$ or $b_2>0$ if $\lambda<0$.\par
%
%
{}For $n=0$, for instance, we get the ground-state energy and wavefunction
\begin{equation}
\begin{split}
  &E_{0,l} = \lambda [2a(2l+d) + 2b_1(4a-2l-d+3) - l(l+d-1)], \\
  &\psi_{0,l}(r) \propto r^l (1+\lambda r^2)^{-a} \exp\left(- \frac{b_1}{1+\lambda r^2} - \frac{b_2}{(1+\lambda
     r^2)^2}\right),
\end{split}
\end{equation}
with the constraints
\begin{equation}
\begin{split}
  A&= 2a(2a+1) + 2b_1(4a-2l-d+3), \\
  B_1 &= 2b_1(4a+2b_1+3) - 4b_2(4a-2l-d+5), \\
  B_2 &= - 4b_1^2 + 4b_2(4a+4b_1+5).
\end{split}
\end{equation}
Furthermore, for $n=1$, we get
\begin{equation}
\begin{split}
  A&= (2a+2)(2a+3) + 2b_1(4a-2l-d+7-4z_1) - 16b_2z_1(z_1-1), \\
  B_1 &= 2b_1(4a+2b_1+7) - 4b_2(4a-2l-d+9-4z_1), \\
  B_2 &= - 4b_1^2 + 4b_2(4a+4b_1+9).
\end{split}
\end{equation}
where $z_1$ is a real solution of
\begin{equation}
  8b_2z_1^3 + 4(b_1-2b_2)z_1^2 - (4a+4b_1+3)z_1 + 4a-2l-d+3 = 0.
\end{equation}
Hence, we may take
\begin{equation}
\begin{split}
  z_1 &= - \frac{b_1-2b_2}{6b_2} + \left(- \frac{v}{2} + \sqrt{\left(\frac{v}{2}\right)^2 + 
    \left(\frac{u}{3}\right)^3}\right)^{1/3} \\
  &\quad {} + \left(- \frac{v}{2} - \sqrt{\left(\frac{v}{2}\right)^2 + \left(\frac{u}{3}\right)^3}\right)^{1/3}, \\
  u &= - \frac{1}{8b_2} \left(4a+4b_1+3 + \frac{2(b_1-2b_2)^2}{3b_2}\right), \\
  v &= \frac{1}{8b_2} \left(4a-2l-d+3 + \frac{2(b_1-2b_2)^3}{27b_2^2} + \frac{(b_1-2b_2)(4a+4b_1+3)}
    {6b_2}\right),
\end{split}
\end{equation}
and we obtain
\begin{equation}
\begin{split}
  E_{1,l} &= \lambda[2a(2l+d+4) + 2b_1(4a-2l-d+7-4z_1) - 16b_2z_1(z_1-1) \\
  &\quad {} - l(l+d-1) + 6], \\ 
  \psi_{1,l}(r) &\propto r^l (1+\lambda r^2)^{-a-1} [1 - z_1(1+\lambda r^2)] \exp\left(- \frac{b_1}{1+\lambda r^2} 
    - \frac{b_2}{(1+\lambda r^2)^2}\right),
\end{split}
\end{equation}
which may correspond to a ground or first-excited state according to the $z_1$ value.\par
%
%
\subsection{Third potential of the second family}

{}For
\begin{align}
  V^{(2)}_3(r) &= \lambda A - \frac{\lambda A}{1+\lambda r^2} + \frac{\lambda B_1}{(1+\lambda r^2)^2} +
    \frac{\lambda B_2}{(1+\lambda r^2)^3} + \frac{\lambda B_3}{(1+\lambda r^2)^4} + \frac{\lambda B_4}
    {(1+\lambda r^2)^5} \nonumber \\
  &\quad {} + \frac{\lambda B_5}{(1+\lambda r^2)^6} + \frac{\lambda B_6}{(1+\lambda r^2)^7}
\end{align}
and the gauge transformation
\begin{equation}
  \tilde{\psi}(z) = z^a \exp(-b_1z - b_2z^2 - b_3z^3) \phi(z),
\end{equation}
the constraints (\ref{eq:epsilon}) and
\begin{equation}
  B_4 = - 16b_2^2 - 24b_3(b_1-2b_2), \qquad B_5 = - 12b_3(4b_2-3b_3), \qquad B_6 = - 36b_3^2,
\end{equation}
expressing $\epsilon$, $B_4$, $B_5$, and $B_6$ in terms of $a$, $b_1$, $b_2$, and $b_3$ (or vice versa), lead to the equation
\begin{align}
  & \biggl\{- 4z(1-z) \frac{d^2}{dz^2} + 2 [-12b_3z^4 -4(2b_2-3b_3)z^3- 4(b_1-2b_2)z^2 + (4a+4b_1+3)z \nonumber \\
  & - 4a + 2l + d -3] \frac{d}{dz} + [B_3 + 16b_2(b_1-b_2) - 6b_3(4a+4b_1+7)]z^3 \nonumber \\
  & + [B_2 + 4b_1^2 - 4b_2(4a+4b_1+5) + 6b_3(4a-2l-d+7)]z^2 \nonumber \\
  & + [B_1 - 2b_1(4a+2b_1+3) + 4b_2(4a-2l-d+5)]z - A + 2a(2a+1)\nonumber \\
  & + 2b_1(4a-2l-d+3)\biggr\} \phi(z) = 0. 
\end{align}
\par
%
%
The Bethe ansatz method yields
\begin{equation}
\begin{split}
  B_3 &= - 16b_2(b_1-b_2) + 6b_3(4a+4n+4b_1+7), \\
  B_2 &= - 4b_1^2 + 4b_2(4a+4n+4b_1+5) + 6b_3\Bigl(4\sum_i z_i - 4a-4n+2l+d-7\Bigr), \\
  B_1 &= 2b_1(4a+4n+2b_1+3) + 4b_2\Bigl(4\sum_i z_i - 4a-4n+2l+d-5\Bigr) \\
  &\quad {}+24b_3\Bigl(\sum_i z_i^2 - \sum_i z_i\Bigr), \\
  A &= (2a+2n)(2a+2n+1) - 2b_1\Bigl(4\sum_i z_i - 4a-4n+2l+d-3\Bigr) \\ 
  &\quad {} - 16b_2\Bigl(\sum_i z_i^2 - \sum_i z_i\Bigr) - 24b_3\Bigl(\sum_i z_i^3 - \sum_i z_i^2\Bigr), 
\end{split}
\end{equation}
with the $z_i$'s solutions of
\begin{align}
  &\sum_{j\ne i} \frac{2}{z_i-z_j} \nonumber \\
  &+ \frac{-12b_3z_i^4 - 4(2b_2-3b_3)z_i^3 - 4(b_1-2b_2)z_i^2 + (4a+4b_1+3)z_i - 4a+2l+d-3}{2z_i(z_i-1)}
    \nonumber \\ 
  &= 0, \qquad i=1, 2, \ldots, n.
\end{align}
We obtain
\begin{align}
  E_{n,l} &= \lambda \Bigl[2a(4n+2l+d) + 2n(2n+1) + 2b_1\Bigl(4a+4n-2l-d+3-4\sum_i z_i\Bigr) \nonumber \\
  &\quad {}- 16b_2\Bigl(\sum_i z_i^2 - \sum_i z_i\Bigr) -24b_3\Bigl(\sum_i z_i^3 - \sum_i z_i^2\Bigr) - l(l+d-1)\Bigr]
\end{align}
and the normalization condition is now $2a>l+\frac{d-1}{2}$ if $\lambda>0$ or $b_3>0$ if $\lambda<0$.\par
%
%
{}For $n=0$, we get the ground-state energy and wavefunction
\begin{equation}
\begin{split}
  &E_{0,l} = \lambda [2a(2l+d) + 2b_1(4a-2l-d+3) - l(l+d-1)], \\
  &\psi_{0,l}(r) \propto r^l (1+\lambda r^2)^{-a} \exp\left(- \frac{b_1}{1+\lambda r^2} - \frac{b_2}{(1+\lambda
     r^2)^2} - \frac{b_3}{(1+\lambda r^2)^3}\right),
\end{split}
\end{equation}
together with the constraints
\begin{equation}
\begin{split}
  A&= 2a(2a+1) + 2b_1(4a-2l-d+3), \\
  B_1 &= 2b_1(4a+2b_1+3) - 4b_2(4a-2l-d+5), \\
  B_2 &= - 4b_1^2 + 4b_2(4a+4b_1+5) - 6b_3(4a-2l-d+7), \\
  B_3 &= - 16b_2(b_1-b_2) + 6b_3(4a+4b_1+7),
\end{split}
\end{equation}
while, for $n=1$, the results read
\begin{equation}
\begin{split}
  E_{1,l} &= \lambda[2a(2l+d+4) + 2b_1(4a-2l-d+7-4z_1) - 16b_2z_1(z_1-1) \\
  &\quad {}  - 24b_3z_1^2(z_1-1)- l(l+d-1) + 6], \\ 
  \psi_{1,l}(r) &\propto r^l (1+\lambda r^2)^{-a-1} [1 - z_1(1+\lambda r^2)] \\
  &\quad {}\times \exp\left(- \frac{b_1}{1+\lambda r^2} 
    - \frac{b_2}{(1+\lambda r^2)^2} - \frac{b_3}{(1+\lambda r^2)^3}\right),
\end{split}
\end{equation}
corresponding to a ground or first-excited state, and the constraints are
\begin{equation}
\begin{split}
  A&= (2a+2)(2a+3) + 2b_1(4a-2l-d+7-4z_1) + 16b_2z_1(1-z_1) \\
  &\quad {} + 24b_3z_1^2(1-z_1), \\
  B_1 &= 2b_1(4a+2b_1+7) - 4b_2(4a-2l-d+9-4z_1) - 24b_3z_1(1-z_1), \\
  B_2 &= - 4b_1^2 + 4b_2(4a+4b_1+9) - 6b_3(4a-2l-d+11-4z_1), \\
  B_3 &= -16b_2(b_1-b_2) + 6b_3(4a+4b_1+11),
\end{split}
\end{equation}
where $z_1$ is a real solution of the equation
\begin{equation}
  12b_3z_1^4 + 4(2b_2-3b_3)z_1^3 + 4(b_1-2b_2)z_1^2 - (4a+4b_1+3)z_1 + 4a-2l-d+3 = 0.
\end{equation}
\par
%
%
\subsection{The one-dimensional case}

To get the one-dimensional results from the $d\ge 2$ ones, we proceed as in Section IIID. The only notable difference is a change in the normalization conditions, which now become $2a>p$ if $\lambda>0$ or $b_m>0$ if $\lambda<0$.\par
%
%
As an example, for the first potential $V^{(2)}_1(x)$, with the constraints
\begin{equation}
  A = 2a(2a+1) + 4b_1(2a+1-p), \qquad B_1 = 2b_1(4a+2b_1+3), \qquad B_2 = - 4b_1^2,
\end{equation}
we obtain from (\ref{eq:results-ter})
\begin{equation}
\begin{split}
  &E_{0,p} = \lambda[2a(2p+1) +4b_1(2a-p+1) - p], \\
  &\psi_{0,p}(x) \propto x^p (1+\lambda x^2)^{-a} \exp\left(-\frac{b_1}{1+\lambda x^2}\right),
\end{split}
\end{equation}
with $2a>p$ if $\lambda>0$ or $b_1>0$ if $\lambda<0$. These energy and wavefunction correspond to a ground state for $p=0$ or to a first-excited state for $p=1$.\par
%
%
In Fig.~5, some examples of extended potentials are plotted for $\lambda>0$, the corresponding wavefunctions being displayed in Fig.~6. Figures 7 and 8 show similar comparisons in the $\lambda<0$ case.\par
%
%
\begin{figure}[h]
\begin{center}
\includegraphics{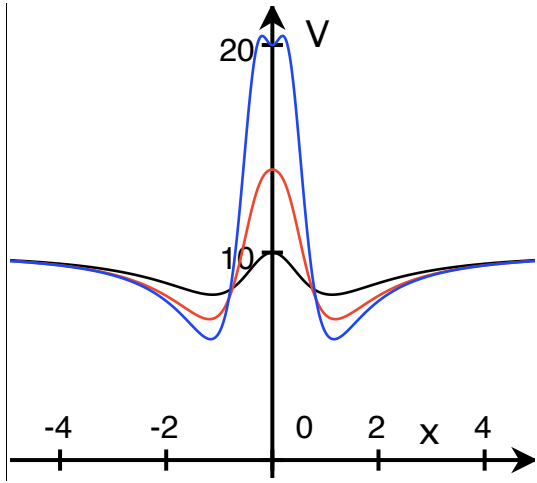}
\caption{Plot of $V^{(2)}_m(x)$ as a function of $x$ for $m=1$, $b_1=1$ (black line), $m=2$, $b_1=b_2=1$ (red line), and $m=3$, $b_1=b_2=b_3=1$ (blue line). In all cases, $\lambda=1$, $a=1/2$, $n=p=0$ (hence $A=10$), and $E_{0,0}=9$.}
\end{center}
\end{figure}
\par
%
%
\begin{figure}[h]
\begin{center}
\includegraphics{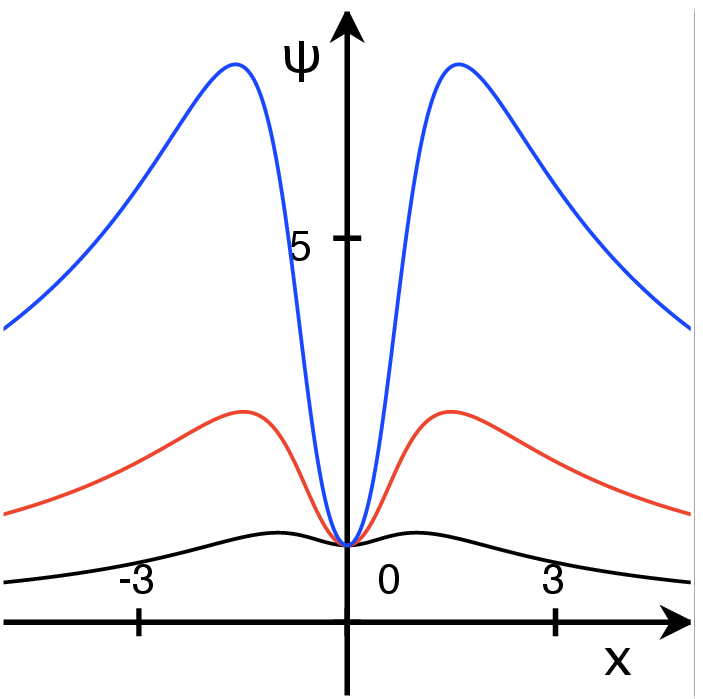}
\caption{Plot of ground-state wavefunction $\psi_{0,0}(x)$ of $V^{(2)}_m(x)$ as a function of $x$ for $m=1$, $b_1=1$ (black line), $m=2$, $b_1=b_2=1$ (red line), and $m=3$, $b_1=b_2=b_3=1$ (blue line). In all cases, $\lambda=1$, $a=1/2$, $n=p=0$ (hence $A=10$), and $E_{0,0}=9$. The wavefunctions are normalized in such a way that $\psi_{0,0}(0)=1$.}
\end{center}
\end{figure}
\par
%
%
\begin{figure}[h]
\begin{center}
\includegraphics{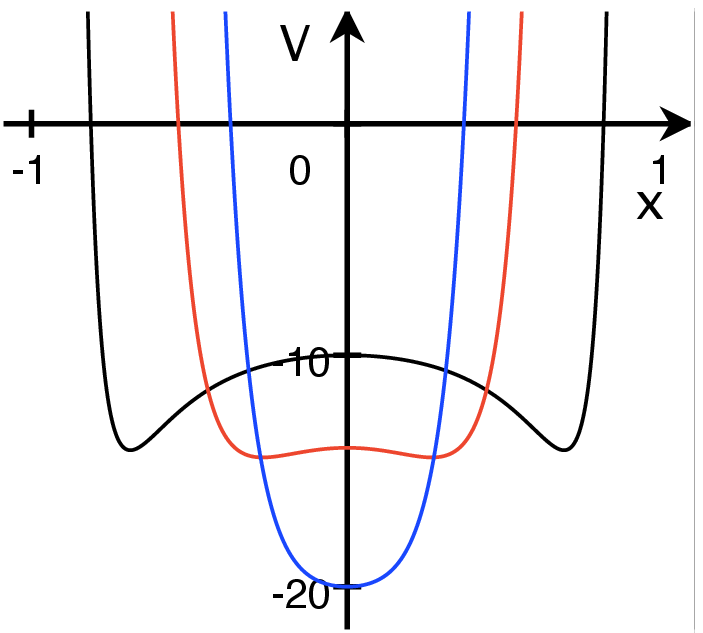}
\caption{Plot of $V^{(2)}_m(x)$ as a function of $x$ for $m=1$, $b_1=1$ (black line), $m=2$, $b_1=b_2=1$ (red line), and $m=3$, $b_1=b_2=b_3=1$ (blue line). In all cases, $\lambda=-1$, $a=1/2$, $n=p=0$ (hence $A=10$), and $E_{0,0}=-9$.}
\end{center}
\end{figure}
\par
%
%
\begin{figure}[h]
\begin{center}
\includegraphics{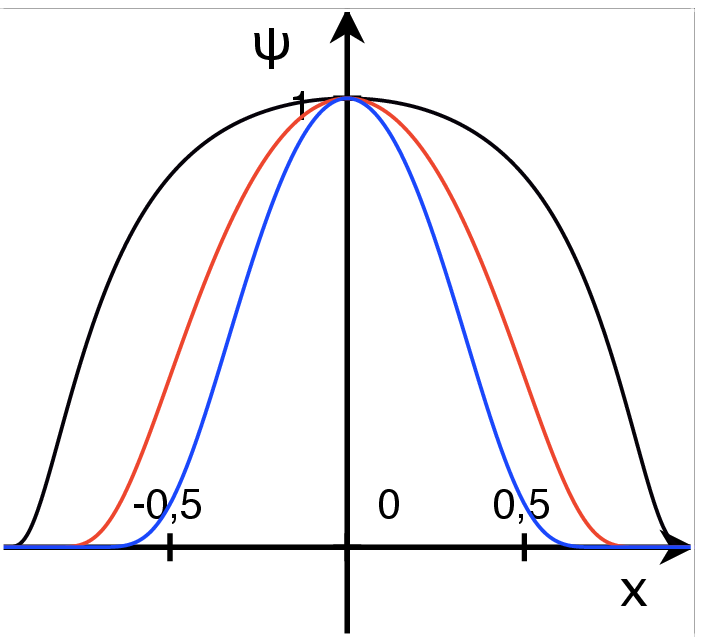}
\caption{Plot of ground-state wavefunction $\psi_{0,0}(x)$ of $V^{(2)}_m(x)$ as a function of $x$ for $m=1$, $b_1=1$ (black line), $m=2$, $b_1=b_2=1$ (red line), and $m=3$, $b_1=b_2=b_3=1$ (blue line). In all cases, $\lambda=-1$, $a=1/2$, $n=p=0$ (hence $A=10$), and $E_{0,0}=-9$. The wavefunctions are normalized in such a way that $\psi_{0,0}(0)=1$.}
\end{center}
\end{figure}
\par
%
%
%
%
\section{CONCLUSION}

In the present paper, we have discussed bound-state solutions to two families of quantum potentials extending the harmonic oscillator in a $d$-dimensional constant-curvature space. We showed that the corresponding radial Schr\"odinger equation for the first three members of both families is reducible to a QES differential equation. Using the Bethe ansatz approach, we obtained closed-form expressions for the energies and wavefunctions of each potential for some allowed values of its parameters. We also proved that the first member of the two families has a hidden sl(2,$\R$) symmetry and is connected with a QES equation of the first or second type, respectively.\par
%
%
{}Furthermore, from the $d$-dimensional radial outcomes, we showed how to derive one-dimensional results valid on the line. In this way, we also obtained two families of QES extensions of the Mathews-Lakshmanan oscillator. It is hoped that our findings will lead to new applications for these nonlinear oscillators.\par
%
%
\section*{APPENDIX: THE FUNCTIONAL BETHE ANSATZ METHOD}

\renewcommand{\theequation}{A.\arabic{equation}}
\setcounter{equation}{0}

Let us consider the second-order differential equation
\begin{equation}
  \left(P(z) \frac{d^2}{dz^2} + Q(z) \frac{d}{dz} + W(z)\right) \phi(z) = 0,  \label{eq:diff-eq}
\end{equation}
where $P(z)$, $Q(z)$, and $W(z)$ are polynomials in $z$ of degree at most $t+2$, $t+1$, and $t$, respectively,
\begin{equation}
  P(z) = \sum_{k=0}^{t+2} p_kz^k, \qquad Q(z) = \sum_{k=0}^{t+1} q_kz^k, \qquad W(z) = \sum_{k=0}^t w_kz^k,
\end{equation}
$p_k$, $q_k$, and $w_k$ being some constants. The functional Bethe ansatz method \cite{zhang, agboola12, agboola13, agboola14} enables one to find under which conditions Eq.~(\ref{eq:diff-eq}) admits polynomial solutions of type (\ref{eq:polynomial}). In the present paper, it will be enough to consider the $t=3$ case.\par
%
%
On substituting $\phi(z) = \prod_{i=1}^n (z-z_i)$ with distinct roots $z_i$ into Eq.~(\ref{eq:diff-eq}), the latter becomes
\begin{equation}
  \sum_{k=0}^5 p_kz^k \sum_{i=1}^n \frac{1}{z-z_i} \sum_{\substack{
        j=1\\
        j\ne i}}^n \frac{2}{z_i-z_j} + \sum_{k=0}^4 q_kz^k \sum_{i=1}^n \frac{1}{z-z_i} + \sum_{k=1}^3 w_kz^k 
        = - w_0. 
\end{equation}
The right-hand side of this equation is a constant, while the left-hand side is a meromorphic function with simple poles at $z=z_i$ and singularity at $z=\infty$. The residue at the simple pole $z=z_i$ is given by
\begin{equation}
  \Res(-w_0)_{z=z_i} = \sum_{k=0}^5 p_k z_i^k \sum_{\substack{
        j=1\\
        j\ne i}}^n \frac{2}{z_i-z_j} + \sum_{k=0}^4 q_k z_i^k. 
\end{equation}
It follows that
\begin{align}
  &\sum_{k=0}^5 p_k \sum_{i=1}^n \frac{z^k-z_i^k}{z-z_i} \sum_{\substack{
        j=1\\
        j\ne i}}^n \frac{2}{z_i-z_j} + \sum_{k=0}^4 q_k \sum_{i=1}^n \frac{z^k-z_i^k}{z-z_i} + \sum_{k=1}^3 w_kz^k
        \nonumber \\
  & +\sum_{i=1}^n \frac{\Res(-w_0)_{z=z_i}}{z-z_i} = - w_0. \label{eq:equation}
\end{align}
\par
%
%
On using the identities
\begin{equation}
\begin{split}
  & \sum_{i=1}^n \sum_{\substack{
        j=1\\
        j\ne i}}^n \frac{1}{z_i-z_j} = 0, \qquad \sum_{i=1}^n \sum_{\substack{
        j=1\\
        j\ne i}}^n \frac{z_i}{z_i-z_j} = \frac{1}{2}n(n-1), \\
   & \sum_{i=1}^n \sum_{\substack{
        j=1\\
        j\ne i}}^n \frac{z_i^2}{z_i-z_j} = (n-1) \sum_{i=1}^n z_i, \qquad \sum_{i=1}^n \sum_{\substack{
        j=1\\
        j\ne i}}^n \frac{z_i^3}{z_i-z_j} = (n-1) \sum_{i=1}^n z_i^2 + \sum_{\substack{
        i,j=1\\
        i<j}}^n z_iz_j, \\
   & \sum_{i=1}^n \sum_{\substack{
        j=1\\
        j\ne i}}^n \frac{z_i^4}{z_i-z_j} = (n-1) \sum_{i=1}^n z_i^3 + \sum_{\substack{
        i,j=1\\
        i\ne j}}^n z_i^2 z_j, 
\end{split}
\end{equation}
Eq.~(\ref{eq:equation}) becomes
\begin{align}
  &[n(n-1)p_5+nq_4+w_3]z^3 + \Bigl\{[2(n-1)p_5+q_4] \sum_i z_i + n(n-1)p_4 + nq_3 + w_2\Bigr\}z^2 \nonumber \\
  &+ \Bigl\{[2(n-1)p_5+q_4] \sum_i z_i^2 + 2p_5 \sum_{i<j} z_iz_j + [2(n-1)p_4+q_3] \sum_i z_i \nonumber \\
  &+ n(n-1)p_3 + nq_2 + w_1\Bigr\}z + [2(n-1)p_5+q_4] \sum_i z_i^3 + 2p_5 \sum_{i\ne j} z_i^2z_j \nonumber \\
  &+ [2(n-1)p_4+q_3] \sum_i z_i^2 + 2p_4 \sum_{i<j} z_iz_j + [2(n-1)p_3+q_2] \sum_i z_i \nonumber \\
  &+ n(n-1)p_2 + nq_1 + \sum_i \frac{\Res(-w_0)_{z=z_i}}{z-z_i} = - w_0. \label{eq:equation-bis}
\end{align}
For the left-hand side of this equation to reduce to a constant, the coefficients of $z^3$, $z^2$, and $z$ must vanish, as well as all the residues at the simple poles. This gives $w_3$, $w_2$, $w_1$ in terms of the coefficients of $P(z)$ and $Q(z)$,
\begin{equation}
\begin{split}
  w_3 &= -n(n-1)p_5 - nq_4, \\
  w_2 &= - [2(n-1)p_5 + q_4] \sum_i z_i - n(n-1)p_4 - nq_3, \\
  w_1 &= - [2(n-1)p_5 + q_4] \sum_i z_i^2 - 2p_5 \sum_{i<j} z_iz_j - [2(n-1)p_4 + q_3] \sum_i z_i \\
  &\quad {} - n(n-1)p_3 - nq_2,
\end{split}
\end{equation}
and the $n$ algebraic equations determining the roots $z_i$,
\begin{equation}
  \sum_{j\ne i} \frac{2}{z_i-z_j} + \frac{\sum_{k=0}^4 q_k z_i^k}{\sum_{k=0}^5 p_k z_i^5} = 0, \qquad i=1, 2, \ldots, n.
\end{equation}
In Eq.~(\ref{eq:equation-bis}), it then remains the constant term leading to
\begin{align}
  w_0 &= - [2(n-1)p_5 + q_4] \sum_i z_i^3 - 2p_5 \sum_{i\ne j} z_i^2 z_j - [2(n-1)p_4 + q_3] \sum_i z_i^2 \nonumber \\
  &\quad {}- 2p_4 \sum_{i<j} z_iz_j - [2(n-1)p_3 + q_2] \sum_i z_i - n(n-1)p_2 - nq_1.
\end{align}
\par
%
%

\end{document}